\documentstyle[12pt,doublespace,epsf]{article}

\begin{document}
\thispagestyle{empty}

\mbox{}
\vspace{1cm}
\begin{center}
{\bf ESCAPE RATES IN BISTABLE SYSTEMS} \\
\vspace{0.5cm}
{\bf INDUCED BY QUASI-MONOCHROMATIC NOISE} \\
\vspace{0.5cm}
{\it S.J.B.Einchcomb and A.J.McKane} \\
\vspace{0.5cm}
 Department of Theoretical Physics \\
University of Manchester \\ Manchester M13 9PL, England \\
\end{center}

\begin{abstract}
Path integral techniques are used to understand the behaviour of a particle
moving in a bistable potential well and acted upon by quasi-monochromatic
external noise.  In the limit of small diffusion coefficient, a steepest
descent
evaluation of the path  integral enables mean first passage times and the
transition times from one well to the other to be computed. The results and
general approach are compared with computer simulations of the process. It is
found that the bandwidth parameter, $\Gamma$, has a critical value above
which particle escape is by white-noise-like outbursts, but below which escape
is by oscillatory type behaviour.
\end{abstract}

\newpage
\pagenumbering{arabic}

\section{Introduction}

The study of noise driven systems frequently focuses on one of the simplest
possible cases, the motion of an overdamped particle in a bistable potential
$V(x)$, in order to investigate the essential features of a particular type
of noise \cite{ref1}.  The model is described by the Langevin equation

\begin{equation} \label{eq:langevin}
\dot{x}+V'(x)=\xi(t)
\end{equation}
where the noise is taken to be Gaussianly distributed with zero mean.  The
nature of the noise is then completely characterised by the correlator

\begin{equation}\label{eq:cor}
\langle \xi(t)\xi(t') \rangle =D~C(|t-t'|)
\end{equation}
Apart from the case of white noise, $C(|t-t'|)=2\delta(t-t')$, the most
studied situation \cite{ref2} is where the noise is exponentially correlated

\[
C(|t-t'|)=\frac{1}{\tau}exp(-|t-t'|/\tau)
\]
The power spectrum

\begin{equation}\label{eq:c}
\tilde{C}(\omega )=\int_{-\infty}^{\infty}~ds~e^{i\omega s}C(|s|)
\end{equation}
may also be used to specify the noise, and the expansion of its inverse

\begin{equation}\label{eq:c-1}
\tilde{C}^{-1}(\omega )=\frac{1}{2}(1+\kappa_{1}\omega^{2}
+\kappa_{2}\omega^{4}+\ldots)
\end{equation}
introduces coefficients $\kappa_{i}$ which define the type of noise under
consideration.  White noise corresponds to $\kappa_{i}=0$ $i\ge1$ and
exponentially correlated noise $\kappa_{i}=0$ $i\ge2$,
$\kappa_{1}=\tau^{2}$. Clearly the most straightforward generalisation
consists in taking $\kappa_{i}=0$ $i\ge3$. The case of a single noise
correlation time and $\kappa_{1},\kappa_{2}>0$ has already
been investigated in an
analogous fashion to the exponentially correlated case \cite{ref3}.
Since $\tilde{C}^{-1}(\omega)$ must be positive for all $\omega$ (for the
Gaussian functional integral over the noise to exist), $\kappa_{2}$ must
be positive. However, $\kappa_{1}$ need not be, and a clearly interesting
limit is where $\tilde{C}^{-1}(\omega)$ is nearly zero for some value of
$\omega$. This corresponds to what has been called quasi-monochromatic noise
(QMN) \cite{mid} or harmonic noise \cite{sgz}, and is the subject of this
paper. These previous studies suggest that QMN differs in an
essential way from white and exponentially correlated noise in that, for
instance, the particle passes over the top of the potential barrier many times
before making a well transition.  One of the aims of this paper will therefore
be to calculate the escape time for the system (\ref{eq:langevin}) with
QMN. We will use path integrals, as in earlier work by Dykman and co-workers
\cite{mid,mid2,mid3},
however our approach will be quite different and, we believe, more
systematic and transparent.

\section{Path Integral Approach}

The correlation function for QMN in frequency space is

\begin{equation}  \label{eq:qmncor}
\langle\xi (\omega )\xi (\omega ')\rangle
=\frac{2D}{(\omega^{2}-\omega_{0}^{2})^{2}+4\Gamma^{2}\omega^{2}}
2\pi\delta (\omega +\omega ')
\end{equation}
If we define a new diffusion coefficient

\begin{equation} \label{eq:d}
\tilde{D}=\frac{D}{\omega_{0}^{4}}
\end{equation}
then $\tilde{C}^{-1}(\omega )$ is of the form (\ref{eq:c-1}) with
$\kappa _{1}=-2\omega_{0}^{-2}(1-2(\Gamma /\omega_{0})^{2})$ and
$\kappa _{2}=\omega_{0}^{-4}$. In the limit $\Gamma\ll\omega_{0}$,
$\tilde{C}(\omega )$ is sharply peaked at the finite
frequency $(\omega_{0}^{2}-2\Gamma^{2})^{1/2} \approx \omega_{0}$,
hence the name ``quasi-monochromatic noise". We will be working within
this limit for the rest of the paper.
In the same way as exponentially correlated noise may also be viewed as
resulting from a stochastic process $\tau\dot{\xi}+\xi=\eta$, where $\eta$ is
Gaussianly distributed white noise with zero mean and strength D, QMN can be
viewed as the result of passing white noise through a harmonic oscillator
filter

\begin{equation} \label{eq:shm}
\ddot{\xi}+2\Gamma\dot{\xi}+\omega_{0}^{2}\xi=\eta
\end{equation}
where the white noise $\eta$ has strength $D$.  We also require that
$\xi$ and $\dot{\xi}$ are both zero as $t\rightarrow\pm\infty$ for
equation (\ref{eq:shm}) to be equivalent to (\ref{eq:qmncor}).\\
We can use (\ref{eq:shm}) to transform from the probability density functional
for white noise to that for $\xi(t)$ \cite{ajm},

\begin{equation} \label{eq:pdfxi1}
P[\xi]={\cal N}exp \left( -\frac{1}{4D} \int_{-\infty}^{\infty} dt \left[
\ddot{\xi}+2\Gamma\dot{\xi}+\omega_{0}^{2}\xi \right]^{2} \right)
\end{equation}
Now we impose that $\xi,~\dot{\xi}\rightarrow0$ as $t\rightarrow\pm\infty$
so that equation (\ref{eq:pdfxi1}) becomes

\begin{equation} \label{eq:pdfxi2}
P[\xi]={\cal N}'exp
\left(  -\frac{\omega_{0}^{4}}{4D} \int_{-\infty}^{\infty} dt \left[
\xi^{2}-\frac{2}{\omega_{0}^{2}}\left( 1-2\frac{\Gamma^{2}}{\omega_{0}^{2}}
\right)
\dot{\xi}^{2}+\frac{1}{\omega_{0}^{4}}\ddot{\xi}^{2} \right] \right)
\end{equation}
Using (\ref{eq:langevin}) and (\ref{eq:d}) this gives the probability
density functional for $x(t)$

\begin{equation} \label{eq:pdf}
P[x]={\cal N}''J[x]exp \left[ -\frac{S[x]}{\tilde{D}} \right]
\end{equation}
where J[x] is a Jacobian factor which is irrelevant to leading order in $D$
and S[x] is the generalised Onsager-Machlup functional given by

\begin{equation} \label{eq:omf}
S[x]=\frac{1}{4} \int_{-\infty}^{\infty}
[\dot{x}+V'(x)]^{2} -\frac{2}{\omega_{0}^{2}}
\left (1-\frac{2\Gamma^{2}}{\omega_{0}^{2}} \right)
 [\ddot{x}+\dot{x}V''(x)]^{2}+
\frac{1}{\omega_{0}^{4}} [x^{(3)}+\ddot{x}V''(x)+\dot{x}^{2}V'''(x)]^{2}
{}~ dt
\end{equation}

Probability distributions, correlation functions, etc can now be obtained by
integrating over functions $x(t)$ with weight given by (\ref{eq:pdf}).
For $D\rightarrow0$ these path integrals may be evaluated by steepest descents,
the paths that dominate the
integrals are the ones for which $\delta S[x]/\delta x=0$.  Using
(\ref{eq:omf}) we obtain a sixth-order non-linear differential equation for the
paths.  If this is multiplied by $\dot{x}$ and integrated with respect to
time we can derive a fifth order non-linear equation
where $x^{(n)}$ is the $n^{th}$ derivative of $x$ with respect to time, given
by

\[
\omega_{0}^{4}(\dot{x}^{2}-V'^{2})+2\omega_{0}^{2} \left(
1-\frac{2\Gamma^{2}}{\omega_{0}^{2}} \right)
(2x^{(3)}\dot{x}-\ddot{x}^{2}+2\dot{x}^{3}V'''-\dot{x}^{2}V''^{2})
\]
\[
+2x^{(5)}\dot{x}-2x^{(4)}\ddot{x}+(x^{(3)})^{2}+10x^{(3)}\dot{x}^{2}V'''
+10\ddot{x}\dot{x}^{3}V''''+2\dot{x}^{5}V'''''
\]
\begin{equation} \label{eq:5ord}
-2x^{(3)}\dot{x}V''^{2}+\ddot{x}^{2}V''^{2}-4\ddot{x}\dot{x}^{2}V''V'''
+\dot{x}^{4}V'''^{2}-2\dot{x}^{4}V''V''''=0
\end{equation}
Even in the limit $\Gamma/ \omega_{0}\rightarrow0$, this is a formidable
looking equation. It seems natural to try to make a perturbative analysis
in the interesting large $\omega_{0}$ limit, which would be somewhat
analogous to an expansion in $\tau^{2}$ for the exponentially correlated
noise problem. But the QMN case is more subtle; the near vanishing of
$\tilde{C}^{-1}(\omega)$ at $\omega \approx \omega_{0}$ gives rise to
rapidly oscillating solutions which have to be carefully extracted from
(\ref{eq:5ord}). In order to make further progress it is useful to have a
specific potential in mind; we choose the much used bistable potential
$V(x)=-x^{2}/2+x^{4}/4$. In the dimensionless units we are using, we are
working in the limits $\Gamma\ll\omega_{0}$ and $\omega_{0}\gg 1$. The
nature of the extremal solution can be glimpsed by studying what happens
near to the turning points $x=-1,0,+1$ where this potential can approximated
by a parabola. Let $x=x_{i}+\rho$ where $x_{i}$ is the turning point and $\rho$
is a small quantity. Taking only the lowest order of powers of $\rho$ we
get the approximate potential to be
$V(\rho)=a_{i}+\frac{1}{2}b_{i}\rho^{2}$ where
$a_{i}=-1/4$ and $b_{i}=2$ if $x_{i}=\pm 1$ and $a_{i}=0$ and $b_{i}=-1$ if
$x_{i}=0$. Equation (\ref{eq:5ord}), when expressed in terms of $\rho$ will
have solutions of the form
$\rho=\sum_{n=1}^{6} {\cal C}_{n}e^{\alpha_{n}t}$ where ${\cal C}_{n}$ and
$\alpha_{n}$ are constants. Substitution into the equation gives

\begin{equation} \label{eq:solut}
\alpha_{n}=\pm b_{i}~\mbox{or}~
\alpha_{n}=\pm (\Gamma \pm i\Omega)
\end{equation}
where $\Omega=\sqrt{\omega_{0}^{2}-\Gamma^{2}}$.
Clearly we can only have $Re(\alpha_{n})>0$ and $Re(\alpha_{n})<0$
solutions separately corresponding to the correct boundary condition
$\rho\rightarrow0$ as $t\rightarrow\pm\infty$. Note that the oscillatory
solutions are independent of the shape of the potential. In the large
$\omega_{0}$ limit, where the form of the potential is unimportant,
the oscillatory solutions of interest to us have precisely this form.
In order to make this more concrete, first look at the case of
the non-linear potential gradient split into N segments between
x=-1 and x=1, where the potential is approximated by N linear segments:

\begin{equation} \label{eq:piece}
V'(x)|_{x_{i}}=(3x_{i}^{2}-1)\left( x-\frac{2x_{i}^{3}}{3x_{i}^{2}-1} \right)
\end{equation}
Making the substitution
\[
x=\frac{2x_{i}^{3}}{3x_{i}^{2}-1}+\rho
\]
the approximate form of the potential gradient is then given by $V'=b_{i}\rho$
where $b_{i}$ denotes the gradient $3x_{i}^{2}-1$ at the point $x_{i}$.
Once again $\rho$ will be the sum of exponentials and
we can say that locally at the point $x_{i}$ the motion of $x$ has the
form:

\begin{equation}   \label{eq:discrete}
x=\left( \frac{2x_{i}^{3}}{3x_{i}^{2}-1}+{\cal C}_{i}^{1}e^{\pm b_{i}t}\right)+
{\cal C}_{i}^{2}e^{\pm\Gamma t}e^{i\Omega t}+
{\cal C}_{i}^{3}e^{\pm\Gamma t}e^{-i\Omega t}
\end{equation}
where here ${\cal C}_{i}^{n}$ are constants at the particular point $x_{i}$ and
are such that $x$ will be continuous.

Substituting this into equation (\ref{eq:5ord}), (\ref{eq:discrete}) can be
shown to be a solution to lowest order in inverse powers of $\omega_{0}$
with $x_{i}^{3}-x_{i}=0$
which suggests that $x(t)$ has the form:

\begin{equation} \label{eq:xform}
x(t)=z_{0}e^{-\Gamma |t-t_{0}|}e^{i \Omega |t-t_{0}|}
+ z^{*}_{0}e^{-\Gamma |t-t_{0}|}e^{-i \Omega |t-t_{0}|}+x_{sm}(t)
\end{equation}
to lowest order in inverse powers of $\omega_{0}$,
where $z_{0}$ is a complex number so that
$x(t)$ is real .
The parameter $t_{0}$ is such that $-\infty <t_{0}<\infty$ and appears
because of the time-translational invariance involved and
can be set to zero without loss of generality. It is the time at which the
amplitude of the oscillations has its greatest value and as a consequence
the likelihood of a transition being made at this time is enhanced.
The smooth
function $x_{sm}(t)$ represents the motion of the centre of oscillation.
We have ignored terms oscillating at higher frequencies since the
magnitude of these components is related to the ratio of the height of the
power spectrum (\ref{eq:qmncor}). For terms oscillating at frequencies $\pm
n\Omega$ this ratio is approximately $\Gamma^{2}/n^{2}\omega_{0}^{2}$.  Since
we are considering the regime $\Gamma/\omega_{0}\ll 1$ this ratio is very
small.  Hence the assumption that terms oscillating at higher
frequenices have very little effect on the motion and can be ignored.

\section{Calculation of the action}

Returning to the case of a general bistable potential $V(x)$, we substitute
the expected form of the solution (\ref{eq:xform}) into the action given by
(\ref{eq:omf}). For (\ref{eq:omf}) to be finite we require that $\dot{x}$
and $\ddot{x}$ to be continuous at $t=0$ which gives
$z_{0}=\frac{1}{2} x_{0}(1-i\frac{\Gamma}{\Omega})$, where $x_{0}$ is
the amplitude of the oscillations at $t=0$. The action consists of three
contributions: one involving the oscillatory terms in (\ref{eq:xform}) only,
one involving cross terms between the oscillatory terms and time derivatives
of $x_{sm}$ and $V'(x)$ and one involving time derivatives of $x_{sm}$ and
$V'(x)$ only. Taking these in turn, the first is evaluated by simply
substituting in the explicit expression given in (\ref{eq:xform}). This
may be facilitated by integrations by parts which are in effect reversing
the steps taken to get from (\ref{eq:pdfxi1}) to (\ref{eq:pdfxi2}). The result
is
$\Gamma x_{0}^{2}$. The second contribution may be evaluated by
integrating by parts to obtain an integrand which has the form of
$(\dot{x}_{sm}+V'(x))$ multiplied by a function of the oscillating terms
in $x$ which vanishes identically. However the discontinuity in $x^{(3)}$
at $t=0$ gives rise to an extra term from this contribution
which is essentially this
discontinuity multiplied by the time derivative of $(\dot{x}_{sm}+V'(x))$
at $t=0$. Using these results and writing out the third contribution in
full gives for equation (\ref{eq:omf})

\[
S[x_{sm}]=\Gamma x_{0}^{2}
-\frac{2\Gamma}{\omega_{0}^{2}}
x_{0}\frac{d}{dt}(\dot{x}_{sm}+V'(x_{sm},t))|_{0}+
\]
\begin{equation} \label{eq:st5}
\frac{1}{4} \int_{-\infty}^{\infty}
(\dot{x}_{sm}+V'(x_{sm},t))^{2} -\left(
\frac{2}{\omega_{0}^{2}}-\frac{4\Gamma^{2}}{\omega_{0}^{4}}\right)
(\ddot{x}_{sm}+\frac{dV'(x_{sm},t)}{dt})^{2}
+\frac{1}{\omega_{0}^{4}}
(x^{(3)}_{sm}+\frac{d^{2}V'(x_{sm},t)}{dt^{2}})^{2} ~dt
\end{equation}
where $V'(x_{sm},t)$ is defined by:

\begin{equation} \label{eq:vandxsm}
V'(x_{sm},t)=V'\left( z_{0}e^{-\Gamma |t|}e^{i \Omega |t|}
+ z^{*}_{0}e^{-\Gamma |t|}e^{-i \Omega |t|}+x_{sm}(t)\right)
\end{equation}
The potential can be written as a sum of a smoothly varying part
and oscillating parts:

\begin{equation} \label{eq:vsmandvosc}
V'(x_{sm},t)=V'_{sm}(x_{sm},t)+\sum_{n\neq 0} V'_{n}(x_{sm},t)
e^{in\Omega t}
\end{equation}
where $V'_{sm}$ and $V'_{n}$ are smooth functions.
We will denote the second term by $V'_{osc}(x_{sm},t)$.  The first term
equals, by integrating over a period,

\begin{equation} \label{eq:vsm}
V'_{sm}(x_{sm},t)=\frac{1}{2\pi}\int_{0}^{2\pi} V'\left(
z_{0}e^{-\Gamma |t|} e^{is}
+ z^{*}_{0}e^{-\Gamma |t|} e^{-is} +x_{sm}(t)\right) ~ds
\end{equation}
We can now integrate equation (\ref{eq:st5}) by parts to give

\[
S[x_{sm}]=\Gamma x_{0}^{2}+
\]
\[
\frac{1}{4} \int_{-\infty}^{\infty}
(\dot{x}_{sm}+V'_{sm}(x_{sm},t))^{2} -\left(
\frac{2}{\omega_{0}^{2}}-\frac{4\Gamma^{2}}{\omega_{0}^{4}}\right)
(\ddot{x}_{sm}+\frac{dV'_{sm}(x_{sm},t)}{dt})^{2}
+\frac{1}{\omega_{0}^{4}}
(x^{(3)}_{sm}+\frac{d^{2}V'_{sm}(x_{sm},t)}{dt^{2}})^{2} dt+
\]
\begin{equation} \label{eq:st6}
\frac{1}{4} \int_{-\infty}^{\infty}
(2(\dot{x}_{sm}+V'_{sm}(x_{sm},t))+V'_{osc}) \left\{
V'_{osc} +\left(
\frac{2}{\omega_{0}^{2}}-\frac{4\Gamma^{2}}{\omega_{0}^{4}}\right)
\frac{d^{2}V'_{osc}(x_{sm},t)}{dt^{2}}
+\frac{1}{\omega_{0}^{4}}
\frac{d^{4}V'_{osc}(x_{sm},t)}{dt^{4}} \right\} dt
\end{equation}
where contributions due to discontinuities at $t=0$ have been omitted since
they are down by powers of $1/\omega_{0}$ on the leading term.

We now show that the third term of (\ref{eq:st6}) is of order $1/\omega_{0}$.
First consider the oscillating term with $n=\pm 1$. The bracket in the third
term of (\ref{eq:st6}) can be approximated by

\[
\left\{
1
-\left( \frac{2}{\omega_{0}^{2}}-\frac{4\Gamma^{2}}{\omega_{0}^{4}}
\right)
\Omega^{2}+\frac{\Omega^{4}}{\omega_{0}^{4}} \right\} \left(
V'_{+1}e^{i\Omega t}+V'_{-1}e^{-i\Omega t} \right)
\approx 0~~+O\left( \frac{1}{\omega_{0}^{2}} \right)
\]

The terms in the potential oscillating at higher frequencies are also of
order $1/\omega_{0}$ by the same argument used at the end of Section 2.

The action is therefore given by

\begin{equation} \label{eq:st7}
S[x_{sm},x_{0})=\Gamma x_{0}^{2} + \frac{1}{4} \int_{-\infty}^{\infty}
(\dot{x}_{sm}+V'_{sm})^{2} dt~~+~~O\left( \frac{1}{\omega_{0}} \right)
\end{equation}
The extremal paths are found from $\delta S/\delta x=0$, or in terms of
the new variables $\{x_{sm}(t),x_{0}\}$, $\delta S/\delta x_{sm}=0$,
where we are not yet varying $x_{0}$. This yields to lowest order in
$\omega_{0}^{-1}$ the equation

\begin{equation} \label{eq:s}
\frac{d}{dt} (\dot{x}_{sm}(t)+V'_{sm}(x_{sm},t))
-(\dot{x}_{sm}(t)+V'_{sm}(x_{sm},t))
\frac{\partial V'_{sm}(x_{sm},t)}{\partial x_{sm}}=0
\end{equation}
For convenience
this can be separated into two first order coupled differential equations which
are given by

\[
\dot{x}_{sm}(t)+V'_{sm}(x_{sm},t)=f(t)
\]
\begin{equation}\label{eq:xandf}
\dot{f}(t)=\frac{\partial V'_{sm}(x_{sm},t)}{\partial x_{sm}}f(t)
\end{equation}
So now the action given by (\ref{eq:st7}) can be written as

\begin{equation}\label{eq:sf}
S=\Gamma x_{0}^{2}+  \frac{1}{4} \int_{-\infty}^{\infty} f^{2}(t)
dt ~~+~~O\left(\frac{1}{\omega_{0}}\right)
\end{equation}
We now introduce a specific potential, the quartic potential
$V(x)=-x^{2}/2+x^{4}/4$, and attempt to calculate the action for a well
transition and the action for a mean first passage. First we have to find
$V'_{sm}$. It is given by

\begin{equation} \label{eq:vsmforbi}
V'_{sm}(x_{sm},t)=x_{sm}^{3}-x_{sm} \left( 1-\frac{3x_{0}^{2}}{2} \left\{
1+\frac{\Gamma^{2}}{\Omega^{2}} \right\} e^{-2\Gamma |t|} \right)
\end{equation}
We can now use (\ref{eq:vsmforbi}) to integrate equations (\ref{eq:xandf}).
In this case there is no easy transformation of variables $\dot{x}(t)=y(x)$ as
used to solve similar equations in \cite{tjn}.
Notice that in the limit $x_{0}\rightarrow0$ the solution of
equations (\ref{eq:xandf}) is that of the extremal path of the white noise
case, that is $\dot{x}_{sm}(t)=-V'_{sm}$ and $f=0$ or
$\dot{x}_{sm}(t)=+V'_{sm}$ and $f=2V'_{sm}$.  The first case is the {\it
downhill} solution and leads to a zero contribution to the action.  This case
also remains a solution even when $x_{0}\neq 0$.  The second case is the {\it
uphill} solution but this no longer remains a solution if $x_{0}\neq 0$ because
of the explicit time dependence of the potential.

In order to keep the white noise solution as the limiting case of
$x_{0}\rightarrow 0$ we must use the boundary conditions,  along with
equations (\ref{eq:xandf}), given by

\[
x_{sm}(-\infty)=-1
\]
\begin{equation}\label{eq:bounds}
x_{sm}(\infty)=0
\end{equation}

We can now use a substitution $s=e^{-|t|}$
in equations (\ref{eq:xandf}) and integrate them
numerically using the COLSYS package \cite{col}.
We have to ensure that $x_{sm}$ and $f$ are continuous at $t=0$ in order for
the original integral (\ref{eq:sf}) to make sense.  Once we have calculated
$f$ we can obtain the action (\ref{eq:sf}) which is a function of the
parameters $x_{0}, \omega_{0}$ and $\Gamma$. In Figure 1 $S$ is plotted
as a function of $x_{0}$ for several values of the bandwidth parameter
$\Gamma$ and a fixed value of $\omega_{0}$. We find that $S$ varies very
little with $\omega_{0}$ as long as this parameter is large enough and in
what follows we fix it to have the value 10.0. It still remains to
minimise $S$ with respect to $x_{0}$ which means we have to determine the
value of $x_{0}$ for which $\partial S/\partial x_{0}=0$ and
$\partial^{2} S/\partial x_{0}^{2}$ is positive.
The value of the action where these criteria hold is denoted by $S^{*}$ and
is calculated numerically. It is then plotted as a function of the
bandwidth parameter $\Gamma$ in Figure 2.  The ratio $S^{*}/\Gamma$ is
plotted as a function of $\Gamma$ in Figure 3.
In the limit $\Gamma\rightarrow 0$ the value $S^{*}/\Gamma$ will be shown
to tend to a finite quantity.  In the work by Dykman et al \cite{mid,mid2,mid3}
the white noise strength has a linear dependence on $\Gamma$ and so this value
is studied to make a comparison between our work and theirs.

There are a number of interesting points in Figures 1-3 which are worthy of
further discussion. In Figure 1 there are two important limits which can
be picked out. The first is the $x_{0}\rightarrow 0$ limit. One sees that
the action tends to 0.25, which is the value found in the white
noise case --- the barrier
height being 0.25 in dimensionless units. The second limit
is when $x_{0}\rightarrow\infty$. The action now tends to
$\Gamma x_{0}^{2}$ and we see that the major contribution of the action
arises due to the energy associated with the oscillatory motion. In Figure
2 we see that for $\Gamma >\approx 0.46$, the minimal action, $S^{*}$,
is just that obtained when the noise is white. It follows that the
the particle escapes from one well to another by white
noise type outbursts.  This means that the action associated with mean first
passages and the action associated with well transitions are equal. On the
other hand for $\Gamma <\approx 0.46$, the minimal action occurs when
$x_{0}\neq 0$. As a consequence we deduce that the
particle escapes from one well to another by an oscillatory type of motion.
This leads to differences between the action associated with a mean first
passage and the action associated with a well transition.
If one plots the envelope of $x$ against $t$ subject to the boundary
condition (\ref{eq:bounds}) then one finds that the particle actually crosses
the potential barrier top many times before $t=+\infty$. Finally, in Figure 3
one sees that
for $\Gamma >\approx 0.46$, the result is as for white noise (having the
constant value $1/4\Gamma$), whereas
for $\Gamma <\approx 0.46$, $S^{*}/\Gamma$ peaks around $\Gamma
\approx 0.05$. The value at $\Gamma=0$ can be calculated analytically and is
found to be $\frac{2}{3}$, as we now show.

As $\Gamma\rightarrow 0$, the paths for $f$ found numerically
around the point where $\partial S/\partial x_{0}=0$ and
$\partial^{2} S/\partial x_{0}^{2}$ is positive
are found to tend to zero for all values of $t$. Setting $f=0$ in
(\ref{eq:xandf}) enables
us to obtain the uphill solution analytically. For $t<0$ one finds that

\begin{equation}\label{eq:xsg0}
x_{sm}^{-2}(t)=2e^{ -2(t-\frac{3x_{0}^{2}}{4\Gamma}\left(
1+\frac{\Gamma^{2}}{\Omega^{2}} \right) e^{2\Gamma t})}
\int_{-\infty}^{t}
e^{2(t'-\frac{3x_{0}^{2}}{4\Gamma}\left(
1+\frac{\Gamma^{2}}{\Omega^{2}} \right) e^{2\Gamma t'})} dt'
\end{equation}
and the action given by (\ref{eq:sf}) becomes:

\begin{equation}\label{eq:sg0}
S=\Gamma x_{0}^{2}
\end{equation}
We can now expand $e^{2\Gamma t} = 1+2\Gamma t + O(\Gamma^{2})$ and
equation (\ref{eq:xsg0}) becomes:

\begin{equation}\label{eq:xsga0}
x_{sm}^{-2}(t)=2 e^{ \frac{3x_{0}^{2}}{4\Gamma}}
e^{-2\left( 1-\frac{3x_{0}^{2}}{2}\right) t + O(\Gamma)}
\int_{-\infty}^{t}
e^{-\frac{3x_{0}^{2}}{4\Gamma}}
e^{2\left( 1-\frac{3x_{0}^{2}}{2}\right) t' + O(\Gamma)} dt'
\end{equation}
This can now be integrated easily to give:

\begin{equation} \label{eq:xsmg0}
x_{sm}^{2}(t)=\left\{
\begin{array}{ll}
1-\frac{3x_{0}^{2}}{2} & x_{0}^{2}<\frac{2}{3}    \\
0                      & x_{0}^{2} \geq \frac{2}{3}
\end{array}
\right.
\end{equation}
It is easy to check that (\ref{eq:xsmg0}) holds also for positive $t$.

The $\Gamma\rightarrow 0$ limit can be made more transparent by proceeding
in a slightly different way. If we begin from the expression

\begin{equation} \label{eq:vee}
V_{sm}(x_{sm}, t) = -\frac{1}{2}\left( 1-\frac{3x_{0}^{2}}{2} \left\{1 +
\frac{\Gamma^{2}}{\Omega^{2}} \right\}
e^{-2\Gamma |t|}\right) x_{sm}^{2} + \frac {1}{4} x_{sm}^{4}
\end{equation}

\noindent we see that if $x_{0}^{2}<\frac{2}{3}$ and $\Gamma \rightarrow
0$, the potential has two minima given by $x_{sm}^{2}(t) = 1
- \frac{3}{2} x_{0}^{2} e^{-2\Gamma |t|}$. However if
$x_{0}^{2}\geq \frac{2}{3}$ the potential
has a single minimum at $x_{sm}=0$.
These two possibilities are illustrated in Figure 4. Since $x_{sm}$ is
independent of time when $\Gamma = 0$, the system is found in minima of
the potential at all times. This can be seen directly from (\ref{eq:xandf}),
where a rescaling of the time by $\Gamma$ shows that when $\Gamma =0$,
$f=0$ and $V'_{sm}=0$ for all finite times. In spite of these
simplifications it is easier to think of the time evolution
of $x_{sm}(t)$ if we imagine $\Gamma$ very small, but non-zero. Suppose,
first of all, that $x_{0}^{2}<\frac{2}{3}$. In the infinitely distant
past $V_{sm}$ has a double-well form with $x_{sm}=-1$. As $t$ increases
then $x_{sm}$ takes the value $-(1 - \frac {3}{2} x_{0}^{2})^{\frac
{1}{2}}$ and remains at this value until large positive times when it
moves back to -1. This clearly violates the boundary condition
$x_{sm}(+\infty)=0$. Now suppose $x_{0}^{2}\geq \frac {2}{3}$.
Once again
$x_{sm}(-\infty)=-1$, but at some large negative time the double wells
merge into a single well at $x_{sm}=0$. At large positive times double
wells develop again, but in the limit $\Gamma =0$ the system remains at
$x_{sm}=0$, which is the correct value as $t \rightarrow \infty$. We
therefore see that we require $x_{0}^{2}\geq \frac{2}{3}$ to
obtain a correct picture when $\Gamma =0$.

To determine $S^{*}/\Gamma$ in the limit $\Gamma\rightarrow 0$ we have
to minimise $S/\Gamma = x_{0}^{2}$ with respect to $x_{0}$ subject to
the condition $x_{0}^{2}\geq \frac {2}{3}$. This leads to the value
$\frac{2}{3}$ for $S^{*}/\Gamma$ in this limit, as claimed.

\section{Computer Simulation of QMN}

Since QMN can be viewed as arising from the second order differential
equation (\ref{eq:shm}) we cannot easily implement the Fox {\it integral}
algorithm \cite{fox} for noise simulation.
There are two possible approaches. The first is to
use an extension of the Sancho algorithm
\cite{san} which makes use of the Box-Muller algorithm \cite{knuth} for the
generation of Gaussianly distributed random numbers.  Then QMN is generated
using a finite difference form of (\ref{eq:shm}) given by:

\begin{equation} \label{eq:findif}
f(t+\Delta t)=2f(t)-f(t-\Delta t)-2\Gamma\Delta t(f(t)-f(t-\Delta t))-
\Delta t^{2}\omega_{0}^{2}f(t)+N(t)
\end{equation}
where $N(t)$ is

\[
N(t)=(-4D\Delta t^{3}\ln(R_{1}))^{\frac{1}{2}}\cos(2\pi R_{2}).
\]
Here $R_{1}$ and $R_{2}$ are random numbers in the range [0,1] and $\Delta t$
is the step length used in the simulation. The second approach is to
consider the system as three coupled differential equations and
to use the work by Mannella and Palleschi \cite{man}. In the notation of
that paper the equations governing the system are written in the form of
three first order differential equations

\[
\dot{x}_{1}=-V'(x_{1})+x_{2}
\]
\[
\dot{x}_{2}=x_{3}
\]
\begin{equation} \label{eq:man1}
\dot{x}_{3}=-2\Gamma x_{3} -\omega_{0}^{2} x_{2} +\sqrt{2D} \xi(t)
\end{equation}
where $x_{1}=x$, $x_{2}=f$, $x_{3}=\dot{f}$ and $\xi(t)$ is Gaussian white
noise of zero mean and mean square value unity.
This system can be simulated by

\[
x_{1}(t+\Delta t)=x_{1}(t)+\Delta t \left(
-V'(x_{1}(t))+x_{2}(t) \right)~~+~~O (\Delta t^{3/2})
\]
\[
x_{2}(t+\Delta t)=x_{2}(t)+\Delta t x_{3}(t)~~+~~O(\Delta t^{3/2})
\]
\begin{equation} \label{eq:man2}
x_{3}(t+\Delta t)=x_{3}(t) +\Delta t^{1/2}\sqrt{2D} M +
\Delta t\left( -2\Gamma x_{3}(t)-\omega_{0}^{2} x_{2}(t) \right)
{}~~+~~O(\Delta t^{3/2})
\end{equation}
where $M$ is a Gaussianly distributed random number of zero mean and mean
square value unity. This
random number can again be generated by the Box-Muller algorithm
and is given by

\begin{equation}
M=(-2\ln(R_{1}))^{\frac{1}{2}}\cos(2\pi R_{2})
\end{equation}
where $R_{1}$ and $R_{2}$ are as before.

With the simulation step length $\Delta t$ small enough
(in our case equal to 0.001) these two approaches are consistent
to fractions of a percent.  Using small $\Delta t$ also eliminates the need to
use any corrector steps as in \cite{man}.
However a full discussion of the merits of
using different algorithms or different step lengths is not included in this
paper.

During the simulations care must be taken to measure both mean first
passage times and the mean well transition times since the theory predicts
different values of the action for these two quantities.
Since we are dealing with small values of $\tilde{D}=D/\omega_{0}^{4}$
then we expect that these characteristic times will have a simple activation
energy type dependence on the associated action.  It is then natural to extend
this to the limit of $\tilde{D}\rightarrow 0$ by plotting the quantity
(where $\tau$ is either the mean first passage time or the mean well
transition time)

\begin{equation}
S(\Gamma,\omega_{0},\tilde{D})=\tilde{D}\ln(\tau)
\end{equation}
against $\tilde{D}$. This will coincide with $S^{*}$ (the quantity
shown as a function of $\Gamma$ in Figure 2)
when $\tilde{D}$
goes to zero.  As in \cite{man2} the action has a near linear dependence on
$\tilde{D}$ and the value of the actions measured for $\omega_{0}=10$
are compared with the theoretical results in Figures 2 and 3.  The
{\it squares} are the actions for mean well transitions and the {\it circles}
are for a mean first passages.  The line connecting the {\it squares} in Figure
3 is to guide the eye and to help show the qualitative features of the
simulation data.

This method is a useful technique since it allows measurement of the action
from simulation data without the need of having knowledge of any prefactors.

\section{Conclusions}

The theoretical results are only correct to order of inverse powers of
$\omega_{0}$ and so for $\omega_{0}=10.0$ these could result to corrections of
order $10\%$.  The errors resulting from the simulations can occur from:

\begin{enumerate}
\item{Using a limited number of ensembles of simulation runs to predict the
ensemble average of typical well transition and first passage times}.
\item{The fact that finite $\tilde{D}$ was used, although these errors are
limited using the analysis used in \cite{man2}}.
\item{Using a finite but small simulation step length, a full discussion of how
this effects measurements is given in \cite{man2}.}
\item{Using a basic finite-step size expansion for the stochastic equations}
\end{enumerate}

However considering all possible errors the theoretical approach and the
computer simulations agree well in the qualitative description of QMN.
Both show that the type of transition between wells in a bistable potential is
dependent on the bandwidth of the noise $\Gamma$.
Both theory and simulations predict that the value of $S^{*}/\Gamma$ has a
finite value of $\approx 0.67$ at $\Gamma=0$ and that this quantity peaks at
$\Gamma \approx 0.05$.

For large values of $\Gamma$ the oscillatory behaviour of the motion is
suppressed and the system behaves as if it is under the influence of white
noise. As $\Gamma$ is reduced however the
particle begins to oscillate and this type of motion causes a well transition,
which results in differences between the action for mean first passage actions
and action for a mean well transition.
Also there is a decrease in size of these two (if
$\tilde{D}$ can be considered $\Gamma$ independent).

This means the bandwidth of
the noise can be used a switching device by the system to decrease the action
and hence escape times between wells.  If a well transition needs to be
inhibited then the bandwidth may be broadened, if it needs boosting then the
bandwidth may be narrowed.  This interesting property and the different method
of motion as compared to white and exponential noise means that many natural
systems may have more control over their behaviour than at first expected.

\section*{Acknowledgements}

SJBE would like to thank the Science and Engineering Research Council (UK) for
a research studentship.

\newpage

\section*{Figure Captions}

\begin{enumerate}
\item{The action calculated from (\ref{eq:sf}) for varying values
of $x_{0}$. Notice how as $\Gamma$ decreases a minimum begins to form
for values of $x_{0}\neq 0$.}
\item{The action $S^{*}$ versus $\Gamma$ together with results from
simulations. {\it Squares} represent
mean well transitions and {\it circles} mean first passages. Both show a
gradual decrease in the value of the action.}
\item{The ratio $S^{*}/\Gamma$ versus $\Gamma$ together with results
from simulations. {\it Squares} represent mean well transitions and
{\it circles} mean first passages. Notice that the value of $S^{*}/\Gamma$
tends to a finite value as
$\Gamma$ tends to zero and that its value peaks at $\Gamma\approx 0.05$.}
\item{The potential $V_{sm}$ (given by (\ref{eq:vee})) versus $x_{sm}$ in
the limit $\Gamma =0$.}
\end{enumerate}

\begin{figure}
\epsfxsize=\textwidth
\epsffile{qmnfig1.eps}
\begin{center}
Figure 1
\end{center}
\end{figure}

\begin{figure}
\epsfxsize=\textwidth
\epsffile{qmnfig2.eps}
\begin{center}
Figure 2
\end{center}
\end{figure}
\begin{figure}
\epsfxsize=\textwidth
\epsffile{qmnfig3.eps}
\begin{center}
Figure 3
\end{center}
\end{figure}
\begin{figure}
\epsfxsize=\textwidth
\epsffile{qmnfig4.eps}
\begin{center}
Figure 4
\end{center}
\end{figure}

\end{document}